# Improving Dysarthric Speech Intelligibility Using Cycle-consistent Adversarial Training

Seung Hee Yang[1], Minhwa Chung[1, 2]
[1]Interdisciplinary Program in Cognitive Science, Seoul National University
[2]Department of Linguistics, Seoul National University, Republic of Korea
sy2358@snu.ac.kr

Abstract: Dysarthria is a motor speech impairment affecting millions of people. Dysarthric speech can be far less intelligible than those of non-dysarthric speakers, causing significant communication difficulties. The goal of our work is to develop a model for dysarthric to healthy speech conversion using Cycle-consistent GAN. Using 18,700 dysarthric and 8,610 healthy control Korean utterances that were recorded for the purpose of automatic recognition of voice keyboard in a previous study, the generator is trained to transform dysarthric to healthy speech in the spectral domain, which is then converted back to speech. Objective evaluation using automatic speech recognition of the generated utterance on a held-out test set shows that the recognition performance is improved compared with the original dysarthic speech after performing adversarial training, as the absolute WER has been lowered by 33.4%. It demonstrates that the proposed GAN-based conversion method is useful for improving dysarthric speech intelligibility.

## 1 INTRODUCTION

Dysarthria refers to a group of speech disorders resulting from disturbances in muscular control of speech mechanism due to a damage in the central or peripheral nervous system. Individuals with dysarthria have problems in verbal communication due to paralysis, weakness, or incoordination of the speech musculature (Darley et al., 1969). It is the most commonly acquired speech disorder affecting 170 per 100,000 persons (Emerson, 1995). Dysarthria is present in approximately 33% of all people with traumatic brain injury, and 20% of persons with cerebral palsy.

Moderate to severe dysarthric speech is often unintelligible to unfamiliar communication partners. To ease the difficulty, speech generating devices in the clinical field of augmentative and alternative communication have been developed (Beukelman and Mirenda, 2005). The devices are developed to recognize and interpret an individual's disordered speech and speak out an equivalent message in a clear synthesized speech. For example, "VIVOCA" commercial device aims to clarify dysarthric speech with an adaptive equalizer in order to amplify the regions of speech that are relevant to speech perception (VIVOCA, 2006). Such technology is also applied in speaking-aid devices that use electrolarynx devices (Bi and Qi, 1997; Nakamura et al., 2006; 2012).

On the other hand, a number of research has been done in speech modification and style conversion towards improving the intelligibility and comprehensibility of dysarthric speech. This is beneficial because dysarthric speakers prefer to use their own natural voicing to express thoughts (Drager et al., 2004). These conversion systems can potentially be used to help people with speech disorders by synthesizing more intelligible or more typical speech (Kain et al., 2007; Hironori et al., 2010; Toda et al., 2012; Yamagishi et al., 2012; Aihara et al., 2013; Tanaka et al., 2013; Toda et al., 2014; Kain and Van Santen, 2009). These transformation systems were used to help people with speech disorders. For example, intelligibility of dysarthric vowels of one speaker was improved from 48% to 54% using an optimal mapping feature set (Kain, 2007). At the short-term spectral level, prosodic extraction and modifications were implemented to improve intelligibility from 68% to 87% (Hosom, 2013).

More recently, Biadsy (2019) introduced an end-to-end-trained speech-to-speech conversion model that maps an input spectrogram directly to another spectrogram. The network is composed of an encoder, spectrogram and phoneme decoders, followed by a vocoder to synthesize a time-domain waveform. The model succeeded in impaired speech normalization. However, the speaker voice identity is lost in this approach. Yang (2019) proposed a GAN-based conversion method that preserves the speaker's voice in the foreign-accented speech domain.

The survey of related works demonstrates that despite the recent growth in spoken language processing technologies, the existing voice conversion methods for dysarthric speech mostly rely on traditional feature engineering methods instead of exploiting the state-of-the-art technologies. In this article, we present our research with the goal of improving intelligibility of dysarthric speech using deep generative modelling. Our approach is to transform the original dysarthric speech signal in the spectral domain by using a cycle-consistent Generative Adversarial Network (CycleGAN) (Zhu et al., 2017), and to synthesize a new speech signal from the trained model. The resulting speech signals are used in an objective evaluation to measure the efficacy of our method.

## 2 SPEECH MATERIAL

Speech database of 100 dysarthric persons of low, moderate and severe degrees of disability has been recorded in a related project called QoLT (Quality of Life Technology) supported by Korea Technology Innovation Program for the purposes of developing technologies to help the disabled lead better lives (Choi, 2011).

### 2.1 Reading Prompts

The reading prompts in Korean comprised of 37 APAC (Assessment of Phonology and Articulation for Children) words, 100 isolated command words, and 50 PBW (Phonetically Balanced Words) words, as shown in Table 1. The command words were designed to represent the Korean alphabet letters for the purpose of speech recognition when controlling electronic appliances, such as a cell phone, PC, TV, and a radio. They corresponded to each Korean alphabet and numbers, analogous to "alpha" for "a" and "bravo" for "b" in English.

In order to avoid words that are difficult to pronounce for people with dysarthria due to complexity or unfamiliarity, 167 nouns with less than four syllables were first extracted from a newspaper corpus (Kim, 2013). Five words were shown to each dysarthric speakers who were asked to select their preferred word for indicating each of the 28 Korean alphabet letters and 10 numbers, in the order of their preference. The list of selected words showed that they preferred food and animal names and shorter words.

Table 1: Composition of Read Speech Tasks and Number of Speakers for Dysarthric and Control Groups in the QoLT database

| Speakers | Reading Prompts |
|---|---|
| 100 Dysarthric Persons | - 37 APAC words<br>- 100 command words<br>- 50 PBW words |
| 30 Control Persons | - 37 APAC words<br>- 100 command words<br>- 150 PBW words |

### 2.2 Dysarthric and Control Speakers

Dysarthric speakers were limited to those with cerebral palsy due to brain damage before or during birth. The speakers suffered from weak limbs and muscles and experience difficulty in articulation due to neurological injury or disease. The age range of the speakers was from 30 to 40, and the ratio of male and female speakers was 2:1. Healthy control speakers were also recorded with the same age and gender distributions. All speakers were recruited from Seoul National Cerebral Palsy Public Welfare. Recordings were made in the environment of a quite office with a Shure SM12A microphone (Choi, 2011).

In order to manually assess the degree of disability, a specialist in speech therapy who had 5 years of experiences in PCC (Percentage of Consonant Correct) assessed the severity of articulation of the dysarthric speakers by listening to the APAC word recordings. For the total 100 dysarthric speakers, four groups were classified from mild (PCC: 85~100%), mild to moderate (PCC: 65~84.9%), moderate to severe; (PCC: 50~64.9%), to severe (PCC: less than 50%) levels, according to diagnostic classification of phonological disorders suggested by Shriberg and Kwiatkowski (Shriberg, 1982). The intra-rater reliability was .957, calculated using Pearson's product moment correlation.

## 3 METHOD

### 3.1 Generative Adversarial Networks

GANs (Goodfellow et al., 2014) have attracted attention for their ability to generate convincing images and speeches. Conditional GANs are able to solve the style transfer problem between black and white images to colored photos, aerial photos to geographical maps, and hand-sketches to product photos, to mention a few (Isola et al., 2017). By interpreting dysarthric speech improvement task as a style transfer problem, we further explore the ability of the generative adversarial learning.

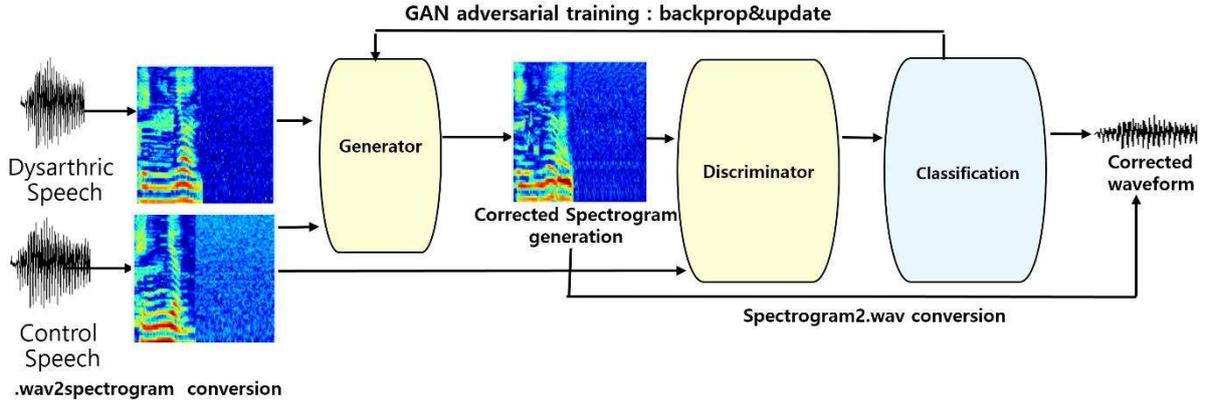

Figure 1. Framework of the proposed model: Control and dysarthric speech are first converted into spectrograms, which are both fed into the generator that outputs fake samples. Discriminator classifies whether the input comes from the generator or the control samples. After training, the generator model is applied to the test spectrograms, and its results are converted back into waveforms for better intelligibility. The figure is adopted from a previous work (Yang and Chung, 2019) that used the framework for accented speech conversion for pronunciation feedback generation.

GANs are generative models that learn to map the training samples to samples with a prior distribution. The generator (G) performs this mapping by imitating the real data distribution to generate fake samples. G learns the mapping by means of an adversarial training, where the discriminator (D) classifies whether the input is a fake G sample generated by G or a real sample. The task for D is to correctly identify the real samples as real, and thereby distinguishing them from the fake samples. The adversarial characteristic is due to the fact that G has to imitate better samples in order to make D misclassify them as real samples.

The misclassification loss is used for further improvement of the generator. During the training process, D back-propagates fake samples from G and correctly classifies them as fake, and in turn, G tries to generate better imitations by adapting its parameters towards the real data distribution in the training data. In this way, D transmits information to G on what is real and what is fake. This adversarial between G and D, which is formulated as:

$$\min_G \max_D V(D,G) = \mathbb{E}_{x \sim P_{data}(x)}[\log D(x)] + \mathbb{E}_{z \sim P_Z(z)}[\log(1-D(G(z)))]. \quad (1)$$

where $P_{data}(x)$ is the real data distribution, and $P_Z(z)$ is the prior distribution. For a given x, $D(X)$ is the probability x is drawn from $P_{data}(x)$, and $D(G(z))$ is the probability that the generated distribution is drawn from $P_Z(z)$.

The adversarial loss alone may not guarantee that the learned function can map the input to the desired output. In our case, this may result in inappropriate or unwanted corrections generated by the network, which is highly undesirable for communication purposes. Cycle-consistency loss was further introduced to reduce the space of possible mapping functions (Zhu et al., 2017). This is incentivized by the idea that the learned mapping should be cycle-consistent, which is trained by the forward and backward cycle-consistency losses described in the following objective function:

$$L_{cyc}(G, F) = \mathbb{E}_{x \sim P_{data}(x)}[\|F(G(x) - x\|_1] + \mathbb{E}_{y \sim P_{data}(y)}[\|G(F(y) - y\|_1]. \quad (2)$$

Here, the network contains two mapping functions $G : X \rightarrow Y$ and $F : Y \rightarrow X$.

This loss fucntion enforces forward-backward consistency between two different domains and is expected to be an effective way to regularize such structured data (Kalal, 2010). Assuming that there is some underlying spectral structure shared between non-native and native linguistic domains, the trained network can be thought of as learning a latent representation of the input that maintains information about the underlying linguistic content. This motivates the proposal in the current study to experiment with CycleGAN.

For each image $x$ from domain X, the translation cycle should be able to bring $x$ back to the original image, and vice versa. While the adversarial loss trains to match the distribution of generated images to the data distribution in the target domain, the cycle-consistency losses can prevent the learned mappings G and F from contradicting each other.

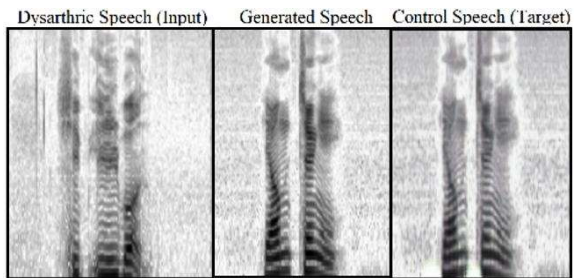

Figure 2. Comparison among input, output, and target spectrograms. The generator learns to imitate the healthy control group's spectrogram by generating a fake version of the reference.

## 3.2 CycleGAN for Improving Dysarthric Speech

In this work, CycleGAN was adopted for the dysarthric to healthy speech mapping task based on two reasons. First, different renderings of the same speech are possible since there can be numerous "golden references," and in theory, there are infinitely many possible outputs. In order to avoid such confusion, cycle-consistency loss seems desirable.

Moreover, despite the differences in the two speech styles, the speech of the same utterance shares an underlying structure. Since cycle-consistency loss preserves the translation cycle, it is our hypothesis that it will have the effect of preserving the global structure of the input spectrograms. Therefore, it is the hypothesis in this work that CycleGAN may have the effect of prohibiting from generating infinitely many possible acceptable outputs while preserving the global structure of the input speech.

In the proposed method, features are automatically learnt in an unsupervised way, in which the generator learns to map dysarthric to control speech distributions. This is implemented in the following in five steps: 1) speech preparation in dysarthric and healthy control domains, 2) speech-to-spectrogram conversion, 3) spectrogram-to-spectrogram training, 4) inversion back to audiosignal, and 5) playback of the generated audio. GAN is used in the third step and the conversion techniques are used during the second and the fourth steps. In order to train using GAN, the prepared samples are fed into the generator, where adversarial training is done using the discriminator which classifies whether the samples are fake (generated speech) or real (healthy speech). The process is shown in Figure 1.

## 3.3 Speech and Spectrogram Conversions

We first convert audio signal to a wideband spectrogram using Short-Time Fourier Transform (STFT) with windows of 512 frames and 33% overlap. It is converted to wideband spectrograms ranging from 100-200Hz, which allows for high quality output in the time resolution. It is then converted to dB amplitude scale, and padded with white noise to generate 128x128 pixels images. White noise padding is done in order to adjust the variability in speech length to a fixed size. We use the python implementation of the Griffin and Lim algorithm to reconstruct the audio signal from the spectrograms by using the magnitude of the STFT (Griffin et al., 1984). The algorithm works to rebuild the signal such that the magnitude part is as close as possible to the spectrogram. For high quality output and minimum loss in transformations, it is run for 1,000 iterations, as suggested in the python implementation of the framework[1].

## 3.4 CycleGAN Implementations

The cycle-consistent adversarial training follows the same network architecture as (Zhu et al., 2017), which adopts the generator architecture from (Johnson et al., 2016) that uses deep convolutional generative adversarial network. For the discriminator training, Markovian PatchGAN (Isola et al., 2017) is used for classifying if each N x N patch in an image is real or fake.

Dysarthric and control spectrogram images are fed into the CycleGAN network. Data augmentation option by flipping images is disabled and batch size was increased to 4 from the default 1. When the training is finished, the model is applied to all the test spectrograms. Web interface visualization of the training process, which was offered in the python implementation, was used to monitor the training and track how spectrogram and the corresponding sounds evolve over time.

## 4 RESULTS AND EVALUATION

Figure 2 shows the spectrograms for dysarthric, generated, and healthy control speeches. It shows that the generator learns to imitate the healthy control group's spectrogram by generating a fake version of the reference. After training, the generator has discovered to generate spectrograms with higher proximity to the health control. Since the test data was completely held out, this means that the model learned to recognize which word the spectrogram represents, and identified how dysarthric speech spectrogram should be mapped to the healthy control spectrogram.

---

[1] https://github.com/bkvogel/griffin_lim

## 4.1 Evaluation

Table 2 displays the automatic speech recognition performance in terms of Word Error Rate (WER) obtained from 100 test samples of generated speech signals and 200 samples of the original signals from the QoLT database. The test samples are generated sounds and contains a certain amount of noise and artifacts, which may lower recognition accuracies. The latter includes clean speeches from control and dysarthric groups. An open source speech recognition engine is used (Google, 2018).

The isolated word recognition of the clean speech of the control group shows 7% error rate. Comparing the recognition accuracies of the original dysarthric with the generated speech the system performance is improved using the proposed method. The absolute WER has been lowered by 33.4%, even with some artifacts introduced in the course of lossy conversions. The performance gain in this objective evaluation suggests that the proposed method can bring benefits to dysarthric speech intelligibility.

Table 2: Automatic speech recognition performances of the proposed CycleGAN approach. Open source API is used and is measured in WER.

| Type | Control Speech | CycleGAN Generated Speech | Dysarthric Speech |
|---|---|---|---|
| WER | 7 | 33.3 | 66.7 |

## 5 CONCLUSION

This study lays the groundwork for an automatic speech correction system for dysarthria speech. Motivated by the visual differences found in spectrograms of dysarthric and control group speeches, we investigated applying GAN to enhance dysarthric speech intelligibility by utilizing generator's ability through adversarial training. Because this mapping is highly under-constrained, we also adopt cycle consistency loss to encourage the output to preserve the global structure, which is shared by dysarthric and control group utterances. Trained on 18,700 spectrogram images in the dysarthric domain and on 8,610 spectrogram images in the control group domain, composed of short utterances in Korean, the generator is able to successfully transform the impaired speech spectrogram input to a spectrogram. The evaluation shows that cycle-consistent adversarial training is a promising approach for dysarthric speech conversion task. In our future work, we plan to conduct perception test in order to verify the improved intelligibility. Also, we plan to improve speaker voice imitability and explore other speech reconstruction method to minimize the noise in the generated speech.

## 6 ACKNOWLEDGEMENT

This research is supported by Ministry of Culture, Sports and Tourism(MCST) and Korea Creative Content Agency(KOCCA) in the Culture Technology(CT) Research & Development Program 2020.